\documentclass[reprint,preprintnumbers,amsmath,amssymb,aps,prd,superscriptaddress,nofootinbib]{revtex4-2}

\usepackage{graphicx}
\usepackage{dcolumn}
\usepackage{bm}
\usepackage{hyperref}
\usepackage[mathlines]{lineno}
\usepackage{xcolor}
\usepackage{booktabs}
\usepackage{enumitem,kantlipsum}

\usepackage{graphicx,subcaption} 

\begin{document}

\preprint{}

\title{Finding Trafficked Radiological Materials via Coherent Elastic \\ Neutrino-Nucleus Scattering}

\author{B.N. Ryan} 
\email{Contact Author: bnryan@mit.edu}
\affiliation{Laboratory of Nuclear Science, Massachusetts Institute of Technology, Cambridge, MA 02139}
\affiliation{Department of Nuclear Science and Engineering, Massachusetts Institute of Technology, Cambridge, MA 02139}

\author{H.D. Pinckney} \affiliation{Laboratory of Nuclear Science, Massachusetts Institute of Technology, Cambridge, MA 02139}

\author{M.P. Short} 
\affiliation{Department of Nuclear Science and Engineering, Massachusetts Institute of Technology, Cambridge, MA 02139}

\author{J.A. Formaggio} \affiliation{Laboratory of Nuclear Science, Massachusetts Institute of Technology, Cambridge, MA 02139}

\date{\today}
             
\begin{abstract}

The potential to use neutrinos for nuclear non-proliferation has been heavily debated due to the tension between production abundance and low interaction rate. A newly detected neutrino interaction channel, coherent elastic neutrino-nucleus scattering (CE$\nu$NS), could potentially end this debate due to its improved cross-section compared to other neutrino interactions. This paper presents a feasibility study for the use of CE$\nu$NS superconducting detectors to find trafficked radiological materials.  To do this, we calculated the minimal activity required for situational detection under ideal conditions, without background, at 95\% confidence level.  This analysis was performed for four commonly smuggled radioisotopes: $^{137}$Cs, $^{109}$Cd, $^{192}$Ir, and $^{57}$Co. Using these results, we conclude that CE$\nu$NS could be used to discover trafficked $^{137}$Cs sources with an activity above the PBq level, but that it is not applicable for finding other radioactive sources.  
This framework can also be applied to other nuclear security concerns, such as safeguarding generation IV nuclear reactors.

\end{abstract}

\maketitle

\section{\label{sec:intro} Introduction}

The potential to utilize neutrinos as a nuclear safeguard asset has been explored for over 40 years \cite{intro}. Their low interaction cross-section makes them extraordinarily impractical to shield \footnote{The average neutrino can go through 50 light-years of lead without interacting. \cite{BrookhavenNeutrino2001}} but, simultaneously, difficult to detect. Most neutrino detection for nuclear safeguards has been focused on $\bar{\nu_e}$-induced inverse beta decay (IBD) as it is the most well-understood anti-neutrino interaction, with a clear detection signal \cite{ibd_signal}.  However, as of 2017 the COHERENT collaboration confirmed a new, promising detection channel called coherent elastic neutrino-nucleus scattering (CE$\nu$NS) \cite{coherent}. CE$\nu$NS is a neutral current interaction that occurs when a neutrino or antineutrino scatters off of a nucleus and exchanges a Z boson, resulting in the nucleus recoiling. In CE$\nu$NS, the neutrino interacts with the entire nucleus due to the low-momentum transfer that occurs.  This is referred to as coherence, and allows for a higher cross-section than achievable with IBD. Coherent interaction cross sections are directly related to the neutron number ($N$) and scale with a factor of $N^2$. This improvement in cross-section allows for smaller detectors and higher count rates, potentially making it a more useful method of detection in certain circumstances. Despite its advantages, CE$\nu$NS has two significant downsides. Firstly, CE$\nu$NS results in a much smaller energy signature (sub-10 keV \cite{recoil}), making detection difficult. Secondly, without knowing the direction of the nucleus’s recoil we are unable to ascertain the energy of the incoming neutrino. 

One aspect of nuclear non-proliferation that could potentially benefit from neutrino detection technology is finding trafficked radiological materials. While radiological materials are not used to make nuclear weapons, they could still be used in dirty bombs and devices.  Some uses of concern are the creation of radiation dispersal devices (RDDs) and radiation exposure devices (REDs).  RDDs are explosives with radioactive materials attached to them.  This results in the dispersal of radioactive material throughout the explosion radius.  While RDDs do not provide a large radiological threat to the population, they would lead to expensive clean-up processes and invoke mass terror \cite{nrc}. REDs are when radioactive sources are planted in places of common public use, such as on subways or in malls \cite{remm}.  Devices such as these can lead to large populations being exposed to increased levels of radiation, making monitoring the legal and illegal transport of these materials and devices crucial to public safety. 

There are two main cases in which we could imagine using a stationary neutrino detector to find trafficked radiological materials.  Firstly, one could use such a detector at border checkpoints, where we define a feasible detection scenario to be a detector located 1-3 m away, detecting for no more than an hour.  Secondly, one could use such a detector at a port, where we define a feasible detection scenario to be a detector located 5-10 m away, detecting for a day. We can further define feasibility by constraining detector mass based on existing dilution refrigerator payloads.

In this study, we focus on superconductors as a possible target detector material. When a particle impacts a superconductor, it creates phonons and quasiparticles (QPs) \cite{qps}. These excitations can then be detected using cryogenic sensors \cite{sensors}. The low energy of the signal quanta (meV scale) leads to a correspondingly low energy threshold when compared to non-superconducting detectors (eV scale). Ultimately, this sub-eV threshold could lead to the detection of keV scale neutrinos, approximately three orders of magnitude lower in threshold than IBD ($\ge 1.8$ MeV \cite{ibd}).

\begin{table*}
    \begin{center}
        \begin{tabular}{llllll}
        \toprule
        Isotope & $^{137}$Cs  & $^{241}$Am \hspace{0.2cm} & $^{109}$Cd & $^{192}$Ir & $^{57}$Co \\
        \midrule
        Incident Rate & 57\% & 48\% & 13\% & 6\% & 6\% \\
        Decay Type \cite{kaeri} & $\beta^-$ & $\alpha$ & EC & $\beta^-$, EC & EC \\
        Case 1 Event & NIS 01/2005 & --- &  CNS 08/2018 & CNS 04/2019 & CNS 03/2015 \\
        Case 1 Activity (Bq) & $2.667 \times 10^{17}$ & ---- & $1.480 \times 10^{9}$ & $5.883 \times 10^{12}$ & $4.440 \times 10^{8}$\\
        Case 2 Event & NIS 04/2012 \hspace{0.2cm} & --- & CNS 01/2013 \hspace{0.2cm} & CNS 09/2018 \hspace{0.2cm} & CNS 09/2022 \\
        Case 2 Activity (Bq) \hspace{0.2cm} & $1.607 \times 10^{15}$ & ---- & $1.240 \times 10^{9}$ & $3.996 \times 10^{12}$ & $3.480 \times 10^{8}$\\
        \bottomrule
        \end{tabular}
    \end{center}
    \caption{The top five most commonly smuggled radioisotopes according to the CNS Global Incidents and Trafficking Database \cite{cns}. In this table, the incident rate is the percent of theft incidents that involved the specified isotope.  Incidents can contain multiple isotopes, and so incident rate totals to greater than $100\%$. }
    \label{tab:isotopes}
\end{table*}

In this paper, we will assess whether or not CE$\nu$NS could be used to find trafficked radiological materials.  To do this, we developed an analysis method to determine the theoretical minimum size of a detector (in units of mass-time). We defined this threshold as the detector mass required to be 95\% confident that there is not a radioactive source being smuggled. Based upon the CNS Global Incidents and Trafficking Database \cite{cns}, we performed this analysis for $^{137}$Cs, $^{109}$Cd, $^{192}$Ir, and $^{57}$Co. Eight example trafficking cases (two for each radioisotope) were considered, using documented incidents from the CNS and NIS Nuclear Trafficking Collection \cite{nis} databases.  This analysis is done assuming ideal detection conditions, which we define as a theoretically low recoil threshold for detection and no background noise.

\section{Isotopes of Concern}\label{sec:isotopes}

 To identify the isotopes and activities of interest, we used the two radiological material trafficking databases available to the general public: the CNS Global Incidents \cite{cns} and Trafficking Database and the NIS Nuclear Trafficking Collection \cite{nis}. The CNS database contains 1558 events from 2013 to 2021, 425 of which were a source being stolen. The top five most commonly stolen radioisotopes from this database are shown in Table \ref{tab:isotopes}. Of the five most stolen isotopes, four of them emit neutrinos in their decay process: $^{137}$Cs \cite{US_Cs137}, $^{109}$Cd, $^{192}$Ir, and $^{57}$Co.  

The NIS database compiles some of the nuclear and radiological trafficking events of concern in former Soviet Union nations between 2005 and 2012. For each of the four isotopes in this study, the two most active sources in the CNS and NIS databases are used here as case studies. Information on these cases can be found in Table \ref{tab:isotopes}. We do note that the $^{137}$Cs cases from the NIS database were more active than those in the CNS database. However, for the feasibility study, we wanted to consider the most active sources between the two databases.

\section{Methods and Results}\label{sec:methods}

To determine whether CE$\nu$NS can feasibly detect trafficked radiological materials, we must calculate the minimum detector exposure to identify the absence of a source to 95$\%$ confidence level (CL). This is done by generating the isotope's neutrino spectrum, the target material's CE$\nu$NS cross section, and combining these to obtain the CE$\nu$NS interaction rate.  Further elaboration on the methods presented here can be found in \cite{my_thesis}. All inputs, scripts, and codes used, along with example outputs, can be found in \cite{rad_mat_trafficking}.

\begin{figure}
    \centering
    \includegraphics[width=1\linewidth]{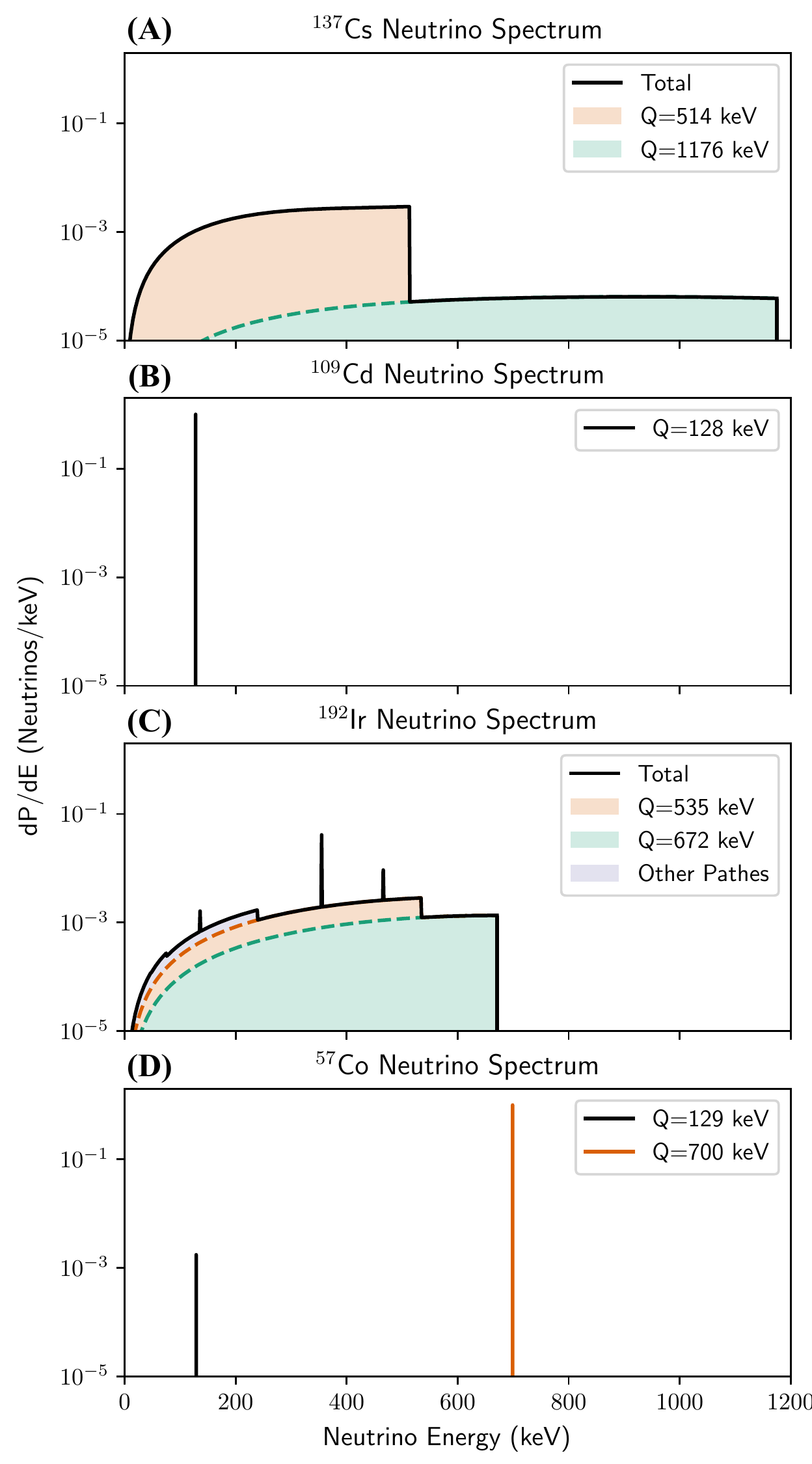}
    \caption{Neutrino spectra for the four isotopes under study: A) $^{137}$Cs, B) $^{109}$Cd, C) $^{192}$Ir, and D) $^{57}$Co. These spectra were generated using \cite{sins}. The delta functions are $\nu_e$ from electron capture decay paths while the continuous curves are $\bar{\nu_e}$ from $\beta^-$ decay. None of the chosen isotopes decayed via $\beta^+$ decay. The different decay paths can be found in \cite{kaeri}.}
    \label{fig:spectra}
\end{figure}

\subsection{Neutrino Spectrum Generation}

Three decay paths can generate neutrinos: $\beta^-$, $\beta^+$, and electron capture (EC). For electron capture, the neutrino energy is equal to the Q-value.  For $\beta$ decay, we calculated the neutrino spectrum using Fermi's theory of beta decay \cite{FermiBeta} with some approximations detailed in \cite{HuberBeta}. The code used to generate the spectra, shown in Fig \ref{fig:spectra}, has been compiled into the SINS (Single Isotope Neutrino Spectrum generator) package \cite{sins}. These spectra are a combination of both $\nu_e$ from EC and $\bar{\nu_e}$ from $\beta^-$ decay. The different decay paths considered can be found in \cite{kaeri}.
 
\subsection{CE$\nu$NS Cross Section Calculation}

The CE$\nu$NS cross-section is a function of the neutrino energy, $E_\nu$, and depends on the recoil threshold energy, $T_0$, and the maximum recoil energy, $T_{max}$.  It is well defined by the standard model, and a simplified version is given by \cite{Formaggio_2012},

\begin{equation} \label{eq:xs}
\sigma(E_\nu)=\frac{G_F^2}{4\pi}E_\nu^2 Q_W^2(1-\frac{T_0}{T_{max}})^2,
\end{equation}
\begin{equation}
    T_{max}=\frac{E_\nu}{1+\frac{M}{2E_\nu}},
\end{equation}

\noindent where $G_F$ is the Fermi constant and $M$ is the mass of the detecting atom. $Q_W$ is the weak nuclear charge, defined as, 

\begin{equation} \label{weak_charge}
Q_W=N-(1-4\sin(\theta_W)^2)Z,
\end{equation} 

\noindent where N is the number of neutrons, Z is the number of protons, and $\theta_W$ is the weak scattering angle.  From the $Q_W$ term, we can see that the CE$\nu$NS cross-section is approximately proportional to $N^2$, as the proportionality to Z is a correction on the order of $O(10^{-2})$.  This motivates using a detector whose target material has a high neutron number.  Eq \ref{eq:xs} is in natural units, where $\hbar=c=1$. Eq \ref{eq:xs} is based on the assumption of a spin-0 nucleus and full coherence (see Appendix C of \cite{my_thesis}).

The theoretical recoil threshold is a property of the detecting material. In the case of superconducting detectors, $T_0$ is the gap energy of the superconductor \cite{ricochet_pres}. Due to its dependence on $T_0$, $M$, and $N^2$, the cross-section for CE$\nu$NS varies widely depending on the detector material used. In this study, we considered superconducting Al, Zn, and Sn to represent a wide range of masses. As $T_0$ differs between these materials, so does the minimum detectable neutrino energy, $E_{\nu, min}$.  Using $T_{max} \geq T_0$ for detection to occur, we define $E_{\nu, min}$ as,

\begin{equation} \label{eq:min_energy}
    E_{\nu, min} = \frac{1}{2}T_0+\frac{1}{2}\sqrt{T_0^2+2T_0M},
\end{equation}

\begin{figure}
    \centering    \includegraphics[width=1\linewidth]{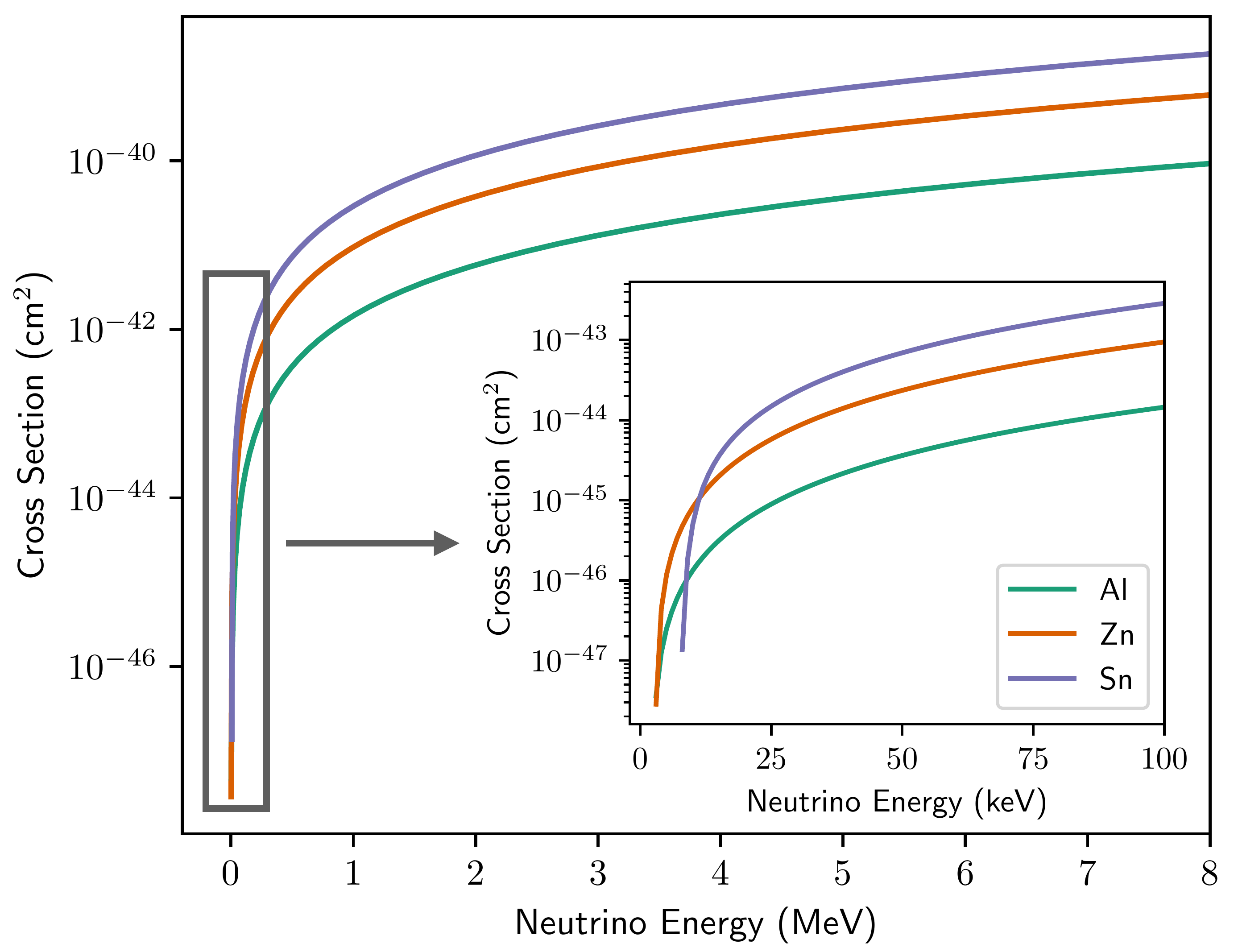}
    \caption{The cross sections of Al, Zn, and Sn, assuming the theoretical minimum recoil threshold, up to 8 MeV. The sub-figure shows cross sections up to 100 keV, highlighting the different minimum-detectable neutrino masses.}
    \label{fig:mat_comp}
\end{figure}

Table \ref{tab:mat_comp} shows the $M$, $T_0$ (calculated using values from \cite{sc_text}), and $E_{\nu, min}$ for Al, Zn, and Sn. These values were used to generate Fig \ref{fig:mat_comp}.  Based on this figure, we determined that Sn has the highest CE$\nu$NS cross section in most situations. The compromise for a higher cross-section is a higher $E_{\nu, min}$ required for detection as well as a lower $T_{max}$, meaning less energy is transferred from the neutrino to the nucleus.

\begin{table}
    \begin{center}
        \begin{tabular}{llll}
        \toprule
        Material & Al & Zn & Sn \\
        \midrule
        M (amu) \cite{kaeri} \hspace{0.2cm} & 27  & 65 & 119 \\
        $T_0$ (meV) & 0.338 \hspace{0.2cm} & 0.241 \hspace{0.2cm} & 1.122 \hspace{0.2cm}\\
        $E_{\nu, min}$ (keV) & 2.062 & 2.711 & 7.877 \\
        \bottomrule
        \end{tabular}
    \end{center}
    \caption{Key parameters of Al, Zn, and Sn as superconducting CE$\nu$NS detectors. The minimum neutrino energy is calculated using Eq \ref{eq:min_energy}.}
    \label{tab:mat_comp}
\end{table}

\subsection{Determining CE$\nu$NS Reaction Rate}

From these cross-sections we can determine the interaction rate as a function of neutrino energy. For small sources, such as smuggled radioactive materials, we approximate the source to be a point source. Based on this assumption, the CE$\nu$NS interaction rate is given by \cite{Formaggio_2012},

\begin{equation} \label{rr-short}
S(E_\nu)=A_s\sigma(E_\nu)\frac{N_A}{A} \frac{M_t}{4\pi \bar{r_i}^2}, 
\end{equation}

\noindent where the variables are defined in Table \ref{tab:rx_rate}. Neutrino oscillation probability does not need to be considered as CE$\nu$NS is flavor blind (which is a property of it being a neutral current interaction).

\begin{table}
    \begin{center}
        \begin{tabular}{lll}
        \toprule
        Variable \hspace{0.2cm} & Definition & Units \\
        \midrule
        $A_s$ & Source Activity & Bq\\
        $N_A$ & Avogadro's Number & mol$^{-1}$\\
        $\rho_t$ & Detector Density & $g/m^3$ \\
        $M_t$ & Detector Mass & g \\
        $V_s$ & Source Volume & $m^3$\\
        $V_{T,i}$ & Differential Target Volume & $m^3$\\
        $r_{st}$ & Source to Target Dis & m\\
        $\bar{r_i}$ & Avg Source to Target Distance \hspace{0.2cm} & m\\
        \bottomrule
        \end{tabular}
    \end{center}
    \caption{Variable definitions and units for Eq \ref{rr-short}.}
    \label{tab:rx_rate}
\end{table}

\subsection{Calculating Minimum Detector Exposure}

We define the minimum exposure as that which is required to confirm the absence of a source to 95\% confidence. As radioactive decay is a Poisson process, we can use Table 40.4 from \cite{pdg}, which gives the confidence interval around the expected number of events ($\lambda$) based on the counts observed in the absence of background. If no counts are observed in a no-background environment, we can be 95$\%$ confident that there is not a source with $\lambda>3.09$ present.

Combining this with Eq \ref{rr-short} we find, 

\begin{equation} \label{eq:important}
    M_{d} = M_t N_t t = \frac{4\pi \lambda A\bar{r_i}^2}{\Bar{\sigma}\Bar{f}(T_0)N_A}\frac{1}{A_s}, 
\end{equation}

\noindent \noindent where $\Bar{\sigma}\Bar{f}(T_0)$ is the average cross-section a neutrino interacting with the detector will have. Individual detector mass ($M_t$), number of detectors ($N_t$), and detecting time ($t$) have been combined into $M_d$, which is a newly defined variable for detector exposure that has units of ton-min.

\begin{figure}
    \centering
    \includegraphics[width=1\linewidth]{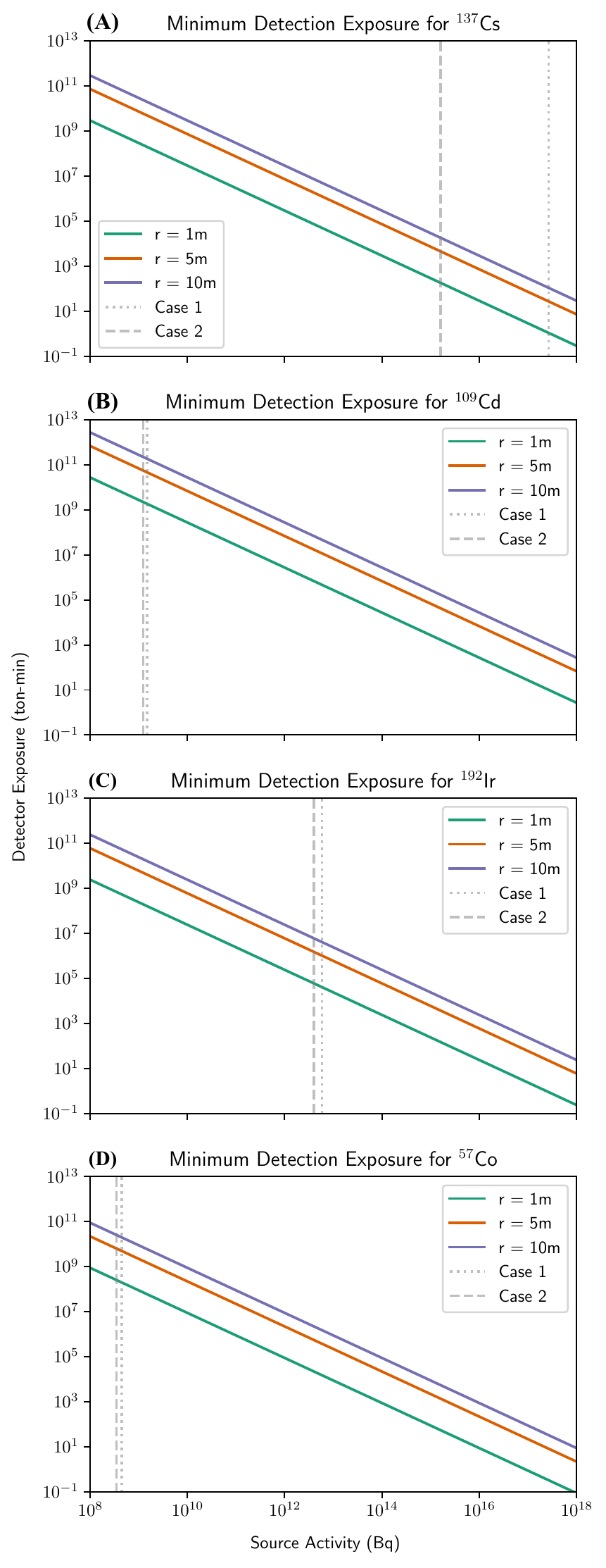}
    \caption{
    Minimum detector exposure for 95\% confidence of source absence as a function of source activity for the following sources: A) $^{137}$Cs, B) $^{109}$Cd, C) $^{192}$Ir, and D) $^{57}$Co. These results were generated using Eq \ref{eq:important}.  Grey dashed lines correspond to cases from the CNS and NIS databases which are detailed in Table \ref{tab:isotopes}.}
    \label{fig:results}
\end{figure}

Figure \ref{fig:results} shows the minimum detector exposure to be 95\% confident that there is not a source with a given activity present at 1$~$m, 5$~$m, and 10$~$m.  For each of these cases, the minimum detector exposure for detection at 1$~$m and 5$~$m was calculated. 

\begin{table*}[]
    \begin{center}
        \begin{tabular}{llllll}
        \toprule
        & Isotope & $^{137}$Cs & $^{109}$Cd & $^{192}$Ir & $^{57}$Co\\
        \midrule
        General Results \hspace{0.2cm} & $\langle\sigma\rangle$ (cm$^2$) & $4.303 \times 10^{-42}$ & $4.726 \times 10^{-43}$ & $5.340 \times 10^{-42}$ & $1.420 \times 10^{-41}$ \\
        & $S$ (per Ci-kg-day) \hspace{0.2cm} & $5.685 \times 10^{-7}$ \hspace{0.2cm} & $7.054 \times 10^{-13}$ \hspace{0.2cm} & $6.598 \times 10^{-7}$ \hspace{0.2cm} & $2.120 \times 10^{-11}$\\
        & Min CUORE $A_s$ (Bq) \hspace{0.2cm} & $6.61\times10^{15}$ \hspace{0.2cm} & $6.02\times10^{16}$ \hspace{0.2cm} & $5.33\times10^{15}$ \hspace{0.2cm} & $2.00\times10^{15}$\\
        & Min Colossus $A_s$ (Bq) \hspace{0.2cm} & $7.40\times10^{14}$  \hspace{0.2cm} & $6.74\times10^{15}$ \hspace{0.2cm} & $5.97\times10^{14}$ \hspace{0.2cm} & $2.24\times10^{14}$ \\
        \midrule
        Case 1 \hspace{0.5cm} & $M_D$ (tons) & 0.017 & $3.05 \times 10^7$ & 679 & $3.39 \times 10^6$\\
        & Feasibility \hspace{0.2cm} & CUORE \hspace{0.2cm} & --- \hspace{0.2cm} & --- \hspace{0.2cm} & ---\\
        \midrule
        Case 2 & $M_D$ (tons) & 3.09 & $3.64 \times 10^7$ & 1000 & $4.32 \times 10^6$\\ 
        & Feasibility & Colossus & --- & --- & ---\\
        \bottomrule
        \end{tabular}
    \end{center}
    \caption{Key results for the isotopes under study and detector mass required to identify a source at a land border. These results were obtained by assuming a Sn detector. S was calculated using Eq \ref{rr-short}. A detector is considered CUORE feasible if it is $\le750$ kg of active mass \cite{cuore} and Colossus feasible if it is $\le6.7$ tons \cite{colossus}. Minimum detectable source activity is also shown for both CUORE and Colossus sized detectors. For $M_D$ required in the ''at port" scenario, simply multiply the results above by $1.042$. }
    \label{tab:discussion}
\end{table*}

\section{Discussion}

\subsection{Feasibility of Detecting Publicly Documented Incidents of Radioactive Material Smuggling}\label{sec:feasibility} 

There are many different metrics and cutoffs that could be used to define detection feasibility. Here, we define feasibility based on the limits of current and near-future cryostats, focusing on the achievable cold mass. This is a limiting factor in superconducting CE$\nu$NS detectors as the detector must operate at extremely low temperatures ($\approx$mK). To our knowledge, CUORE is the largest cryogenic detector in operation, and can cool a total of 4.5 tons, 750 kg of which is active mass \cite{cuore}. Assuming that a CE$\nu$NS superconducting detector would have a similar structural to payload ratio, we adopt 750 kg as our standard for what current dilution fridge technology can achieve and consider this ``CUORE feasible". 

We take the Colossus cryostat, currently under construction, to set the scale of overall feasibility \cite{colossus}.  This cryostat is designed to support a total of 40 tons. Assuming a similar packing ratio to CUORE results in an approximately 6.7 ton active mass payload. As this dilution refrigerator is still under construction and therefore not proven to work, we define detector masses under 6.7 tons as ``Colossus Feasible".

Our standards for exposure time and distance are based on two scenarios: ports and land borders.  For detecting materials at a land border we assume the source is 1m away and our exposure time is an hour. At a port we assume the source is 5m away and our exposure time is a day.  Using these combinations we estimate the total mass required to detect a variety of documented scenarios, detailed in Table \ref{tab:discussion}.  Utilizing our ``CUORE feasible" metric we could detect case 1 of $^{137}$Cs, and utilizing the ``Colossus feasible" metric we can extend this to case 2.  No other cases are detectable using these technologies. 

This conclusion is based on the trafficking cases selected and is therefore database dependent.  Access to a wider variety of radiological material trafficking events, potentially with stronger sources, could change this result. The CNS database is only a small fraction of the total number of radiological material smuggling events, with the International Atomic Energy Association's (IAEA) confidential Incident and Trafficking Database (ITDB) containing 4243 different events, 350 of which are known to have been taken with malicious intent and 1045 that are potentially related to trafficking \cite{ITDB}. Even that database is likely to only have a small fraction of events, 14755 radiation sources were collected in Ukraine from 2009 to 2015, most of which are likely not in the ITDB \cite{ukraine}. Private incident databases could contain higher activity sources, which could potentially be detected using this technology.

\subsection{Currently Operating CE$\nu$NS Detector Case Study}

The work described above was done using theoretically optimal detection parameters consisting of a theoretically low threshold limit and no background noise. However, as CE$\nu$NS is a rather new detection method, we are currently far from these optimal detection parameters. While there is not currently an active superconducting CE$\nu$NS detector, we can estimate some detection parameters from the Ricochet experiment to get an approximation of the current feasibility of finding trafficked radioactive materials with CE$\nu$NS. Ricochet is an array of low-temperature bolometers aiming to detect CE$\nu$NS from the Institut Laue–Langevin (ILL) research reactor in Grenoble, France \cite{AUGIER2023168765}. The next payload to be installed is the superconducting ``Q-Array" with a projected recoil threshold of 50 eV \cite{50ev}. While background levels are yet to be released, we will approximate a $50\%$ detection efficiency due to muon dead time.

\begin{figure}
    \centering
    \includegraphics[width=1\linewidth]{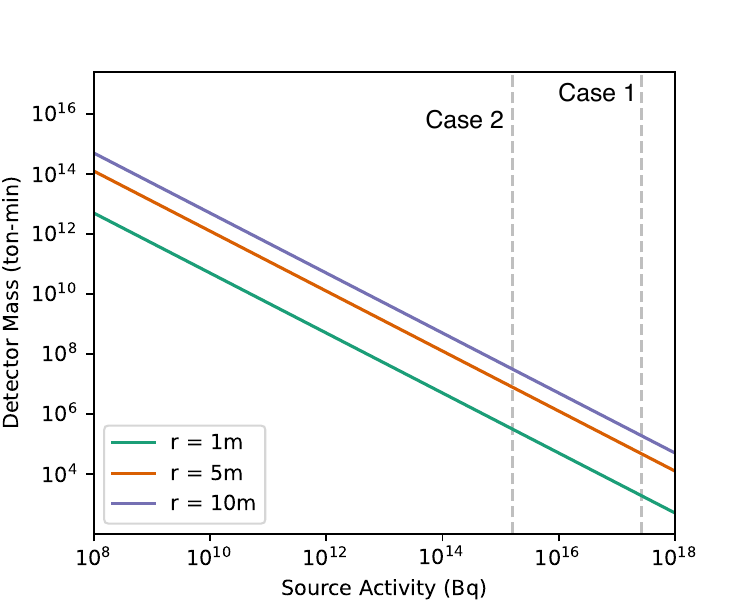}
    \caption{Minimum detector exposure for 95\% confidence of $^{137}$Cs source absence as a function of activity, based on current experimental detection parameters.}
    \label{fig:exp_results}
\end{figure}

Based on this 50 eV recoil threshold, we calculate the minimum neutrino energy required for detection for Al, Zn, and Sn, shown in Table \ref{tab:exp_mat}.  With these minimum neutrino energies, the best material to detect antineutrinos from $^{137}$Cs is superconducting aluminum, as the maximum neutrino energy emitted by $^{137}$Cs is 1.176 MeV \cite{kaeri}.  Using these experimental detection parameters and superconducting aluminum, we performed the same analysis used to obtain Fig \ref{fig:results} and Table \ref{tab:discussion}.  The results are shown in Figure \ref{fig:exp_results} and Table \ref{tab:exp_results}. Based on this assumed detector performance the minimum exposure required to detect case 1's source is $\approx$31.5 tons in both the border and port scenarios. Using the same feasibility standards set in Sec \ref{sec:feasibility}, we conclude that detecting trafficked $^{137}$Cs with current CE$\nu$NS detector parameters is not feasible.

\begin{table}
    \begin{center}
        \begin{tabular}{lll}
        \toprule
        Material \hspace{0.2cm} & Theory $E_{min}$ \hspace{0.2cm} & Exp $E_{min}$ \\
        \midrule
        Al & 2.062 keV & 0.793 MeV \\
        Zn & 2.711 keV & 1.234 MeV \\
        Sn & 7.877 keV & 1.662 MeV \\
        \bottomrule
        \end{tabular}
    \end{center}
    \caption{Lowest detectable neutrino energy for theoretically optimal detectors (Theory $E_{min}$) versus the ricochet projected performance (Exp $E_{min}$).}
    \label{tab:exp_mat}
\end{table}

\begin{table}[]
    \begin{center}
        \begin{tabular}{ll}
        \toprule
        Isotope & $^{137}$Cs\\
        \midrule
        $\langle\sigma\rangle$ (cm$^2$) & $2.124 \times 10^{-43}$ \\
        $S$ (per Ci-kg-day) \hspace{0.2cm} & $1.0578 \times 10^{-7}$ \\
        \midrule
        Case 1 Event & NIS 01/2005 \\
        @ 1m (ton-min) & 1888 \\
        @ 5m (ton-min) & 47205 \\
        \midrule
        Case 2 Event & NIS 04/2012 \\
        @ 1m (ton-min) & $3.134 \times 10^{5}$ \\
        @ 5m (ton-min) & $7.834 \times 10^6$ \\
        \bottomrule
        \end{tabular}
    \end{center}
    \caption{Key results for $^{137}$Cs detection using Ricochet Q-Array experimental parameters \cite{50ev}. These results were obtained assuming an Al detector.}
    \label{tab:exp_results}
\end{table}

\section{Conclusion}

In this study, we assessed the feasibility of detecting trafficked radiological materials based on incidents of source theft and smuggling. To do this, we calculated the minimum detector exposure required to be 95\% confident that, if there were no neutrinos detected, a source of activity X or greater was not present. This analysis assumed optimal detection parameters such as no background and a recoil threshold equal to the gap energy of the superconductor. 

We performed this analysis for $^{137}$Cs, $^{109}$Cd, $^{192}$Ir, and $^{57}$Co, as they are the most commonly smuggled isotopes that emit neutrinos according to the CNS Global Incidents and Trafficking Database.  To assess feasibility we considered two scenarios: detecting materials at a border (1m away, detecting for 1 hour) and at a port (5m away, detecting for 24 hours). We found that it would be feasible with CUORE ($M_D \leq$ 750 kg) to detect the largest $^{137}$Cs source in the NIS database and feasible with Colossus ($M_d \leq$ 6.7 tons) to detect the second largest $^{137}$Cs source.  No $^{109}$Cd, $^{192}$Ir, or $^{57}$Co sources in the CNS or NIS databases would be feasibly detectable in these scenarios. $^{137}$Cs sources with an activity $\ge 7.40\times10^{14}$ Bq are feasible to detect using a Colossus-sized dilution refrigerator.

While we identified that, theoretically, you could detect trafficked $^{137}$Cs with one superconducting tin CE$\nu$NS detector, this result is very database dependent.  In the future, it could be worthwhile to perform the analysis in this paper on more comprehensive radiological trafficking databases that are not available to the general public. The minimum detectable activity, on the other hand, is not database dependent.

Experimentally, there is still a long way to go in detector design before we are close to these optimal detection parameters.  Assuming experimental detection parameters similar to those of Ricochet Q-Array, it is not feasible to detect even the largest $^{137}$Cs source. With a Colossus-sized dilution fridge, it would take 4.7 hours of detection in our border scenario to establish 95$\%$ confidence that none of the examined sources are present. 

The main result of this study was the development of an analysis method to determine the minimum detector mass required to detect specific sources.  While in this paper this was done for single isotopes, in the future it can be applied to more complex neutrino sources such as nuclear reactors.  Future studies could apply this method to other nuclear safeguards concerns such as monitoring new generation 4 reactors, research reactors with a power output of 10 MW$_{th}$ or greater, and dry cask short-term spent nuclear fuel storage \cite{nu_tools}.

\begin{acknowledgments}

This work was supported by the National Science Foundation under grant number 2411390. B.N.R. would also like to thank the Lemelson family for their financial support of this work through the MIT Lemelson Engineering Presidential Graduate Fellowship.

\end{acknowledgments}

\bibliography{june_12_2023}

\end{document}